\documentclass[reprint,amsmath,amssymb,aps]{revtex4-1}
\usepackage{graphicx}
\usepackage{dcolumn}
\usepackage{bm}

\newcommand{\begineq}[1]{\begin{equation}\label{#1}}
\newcommand{\eqend}{\end{equation}}

\newcommand{\reffig}[1]{Fig.\ \ref{#1}}

\newcommand{\jump}[1]{\left[ \! \left[ {#1} \right] \! \right]}

\begin{document}

\preprint{APS/123-QED}

\title{Giant Enhancement of the Electromechanical Coupling in Soft Heterogeneous Dielectrics}
\author{Stephan Rudykh}%
\author{Arnon Lewinstein}%
\author{Gil Uner}%
\author{Gal de{B}otton}
\altaffiliation[Also at ]{Department of Biomedical Engineering, Ben-Gurion University}
\affiliation{%
 Department of Mechanical Engineering, Ben-Gurion University, Beer-Sheva 84105, Israel}

\date{\today}

\begin{abstract}
Electroactive soft elastomers require huge electric field for a meaningful actuation. 
We demonstrate that this can be dramatically reduced and giant deformations can be produced by application of suitably chosen heterogeneous actuators. 
The mechanism by which the enhancement is achieved is described and illustrated with the aid of both idealized and periodic models.
\begin{description}
\item[PACS numbers]
May be entered using the \verb+\pacs{#1}+ command.
\end{description}
\end{abstract}

\pacs{Valid PACS appear here}
\maketitle

\begin{figure}[b]
\includegraphics[scale=0.40]{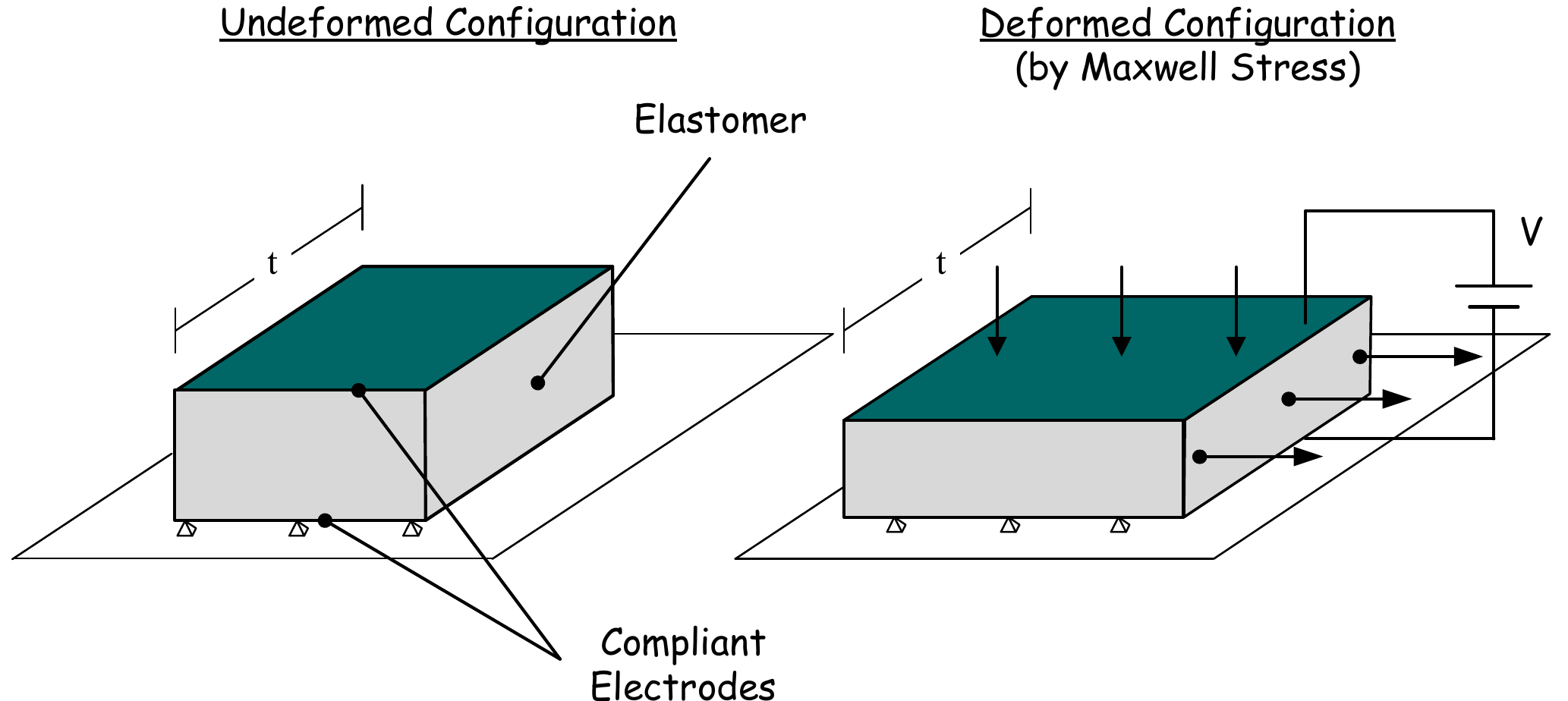}
\caption{A sketch of a planar EAP actuator.}\label{actuator}
\end{figure}

Electroactive polymers (EAP) are capable of large deformations in response to electric stimulus. 
A sketch of a planar actuator is shown in \reffig{actuator}. 
The top and bottom faces of the soft dielectric are covered with compliant electrodes \citep{bhatetal01} inducing an electric field through the material. 
The resulting Maxwell stress leads to the deformation of the material.
The variety of possible applications of these ``artificial muscles'' motivated an intensive search for appropriate polymers. 
Indeed, recent experimental studies achieved remarkable milestones in terms of the magnitudes of the actuation strains
\cite{pelretal00scie,bhatetal01,lacoetal04saa,carpdero05itdei,Carpietal07sms,ohaletal08jap,stoyetal11sm}. 
In parallel, the concept of enhancing the responsiveness of EAP devices by means of snap-through unstable mechanisms was examined too \citep{zhaoetal10prl,mockgoul06ijnm,rudyetal2011ijnm}.
However, these studies did not tackle the main limitation of EAPs, namely the huge electric fields needed for meaningful actuations.

We address this challenge and investigate a mechanism by which the exciting electric field can be reduced by an order of magnitude.
In this regard we recall that recent experimental works \cite{huanetal04aple,stoyetal11sm} involving soft elastomers with high dielectric particles demonstrate an improved response of the heterogeneous systems.
In agreement, theoretical studies \cite{gdbetal07mams,tianetal10submitted} of idealized heterogeneous dielectrics predicted an enhancement of the electromechanical coupling.
In this work the non-linear theory of electroelasticity at \emph{finite strains} \cite{toup56arma,mcmeland05jamt,dorfogde05acmc,suoetal08jmps} is adopted and an exact analytical solution for an idealized heterogeneous system is deduced. 
This, in turn, sheds light on the mechanism that leads to the improved coupling and motivates an investigation of more realistic microstructures.
By application of the finite element (FE) method corresponding periodic models are examined and the improvement in the electromechanical coupling is quantified.

The deformation of the material is characterized by the deformation gradient $F_{ij}={\partial x_{i}}/{\partial X_{j}}$, where $x_{i}$ and $X_{j}$ are the position vectors of a material point in the deformed and reference states, respectively. 
The electric field at a point is $E_{i}=-{\partial\phi}/{\partial x_{i}}$, where $\phi$ is the electric potential.
In incompressible and isotropic neo-Hookean dielectrics
the electric displacement and the total stress tensor are 
\begin{equation}\label{nH}
    D_{i}=\epsilon_{0}\epsilon E_{i}
    \quad\text{and}\quad
    \sigma_{ij}=\mu F_{ik}F_{jk}+ \epsilon_{0}\epsilon E_{i}E_{j}-p\delta_{ij},
\end{equation}
where $\epsilon_{0}$ is the vacuum permeability, $\epsilon$ is the dielectric constant, $\mu$ is the shear modulus, $p$ is the pressure, and $\delta_{ij}$ is the Kronecker delta. 
In the expression for the stress the first term is the mechanical stress and the electrostatic Maxwell stress tensor is the second term.
\begin{figure}[t]
\includegraphics[scale=0.33]{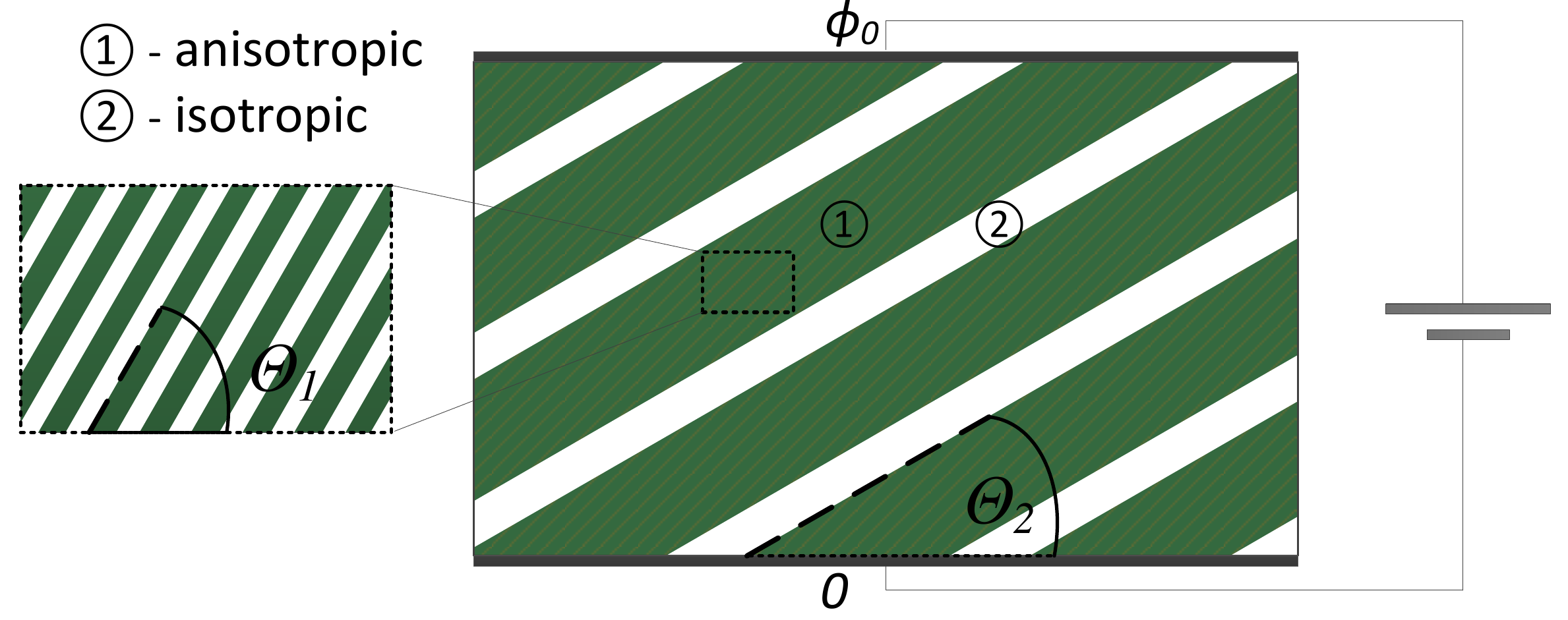}
\caption{A sketch of the idealized layered model.}\label{Lam}
\end{figure}
Assuming a quasistatic deformation, no magnetic fields and no body forces, the governing equations are
\begin{equation}\label{GovEq}
    \dfrac{\partial D_{i}}{\partial x_{i}}=0
    \quad\text{and}\quad 
    \dfrac{\partial\sigma_{ij}}{\partial x_{j}}=0.
\end{equation}
In the absence of free charges at the interfaces the electric field continuity conditions are
$\jump{E_{i}}\hat{m}_{i}=0$ and $\jump{D_{i}}\hat{n}_{i}=0$,
where $\jump{\bullet}\equiv(\bullet)^{+}-(\bullet)^{-}$ is the jump across the interface and
$\hat{n}_{i}$ and $\hat{m}_{i}$ are the unit vectors normal and tangent to the interface, respectively.
The corresponding mechanical continuity conditions are
$\jump{F_{ij}}\hat{m}_{j}=0$ and $\jump{\sigma_{ij}}\hat{n}_{j}=0$.

Figure~\ref{Lam} depicts an idealized layered material with alternating soft isotropic and anisotropic layers (phases 2 and 1).
The anisotropic layers are themselves laminated structures made out of alternating ``sublayers'' of a stiff material with high dielectric modulus and a soft material with low dielectric modulus.
Four parameters characterize this microstructure: the lamination angle with respect to the electrodes plane and the volume fraction of the isotropic layers ($\Theta_{2}$ and $\alpha_{2}$), and the lamination angle and volume fraction of the stiff sublayers in the anisotropic layer ($\Theta_{1}$ and $\alpha_{1}$).
It is assumed that the thickness of the sublayers is an order of magnitude smaller than the thickness of the enclosing layer.  
This ``scale separation'' allows to solve analytically the associated coupled homogenization problem \citep{gdb05jmps,ltd08phd}, and to determine the planar mechanical response to electrostatic excitation $E_{0}=\phi_{0}/d$, where $\phi_{0}$ is the electric potential between the electrodes and $d$ is the distance between them in the undeformed state. 
The parameters of the two models examined in this work are summarized in Table~\ref{Gconstants}.
Model I1 follows an optimization process in the limit of \emph{infinitesimal deformation} that resulted in an extremely thin isotropic layers \cite{tianetal10submitted}. 
Model I2, with thicker isotropic layers, is reminiscent of the periodic models that were inspired by the predictions of the optimized one.
\begin{table}[b]
\caption{\label{Gconstants}%
Microstructure parameters}
\begin{ruledtabular}
\begin{tabular}{ccccc}
 Model &$\alpha_{1}$&\multicolumn{1}{c}{\textrm{$\alpha_{2}$}}&\multicolumn{1}{c}{\textrm{$\Theta_{1}$}}&\multicolumn{1}{c}{\textrm{$\Theta_{2}$}}\\
 \hline
I1 & $0.416$ &$0.008$ & $63.1^{\circ}$ & $27.5^{\circ}$ \\
I2  & $0.6$ &$0.06$    & $63.1^{\circ}$ & $27.5^{\circ}$\\
\end{tabular}
\end{ruledtabular}
\end{table}

\begin{figure}[t]
\includegraphics[scale=0.48]{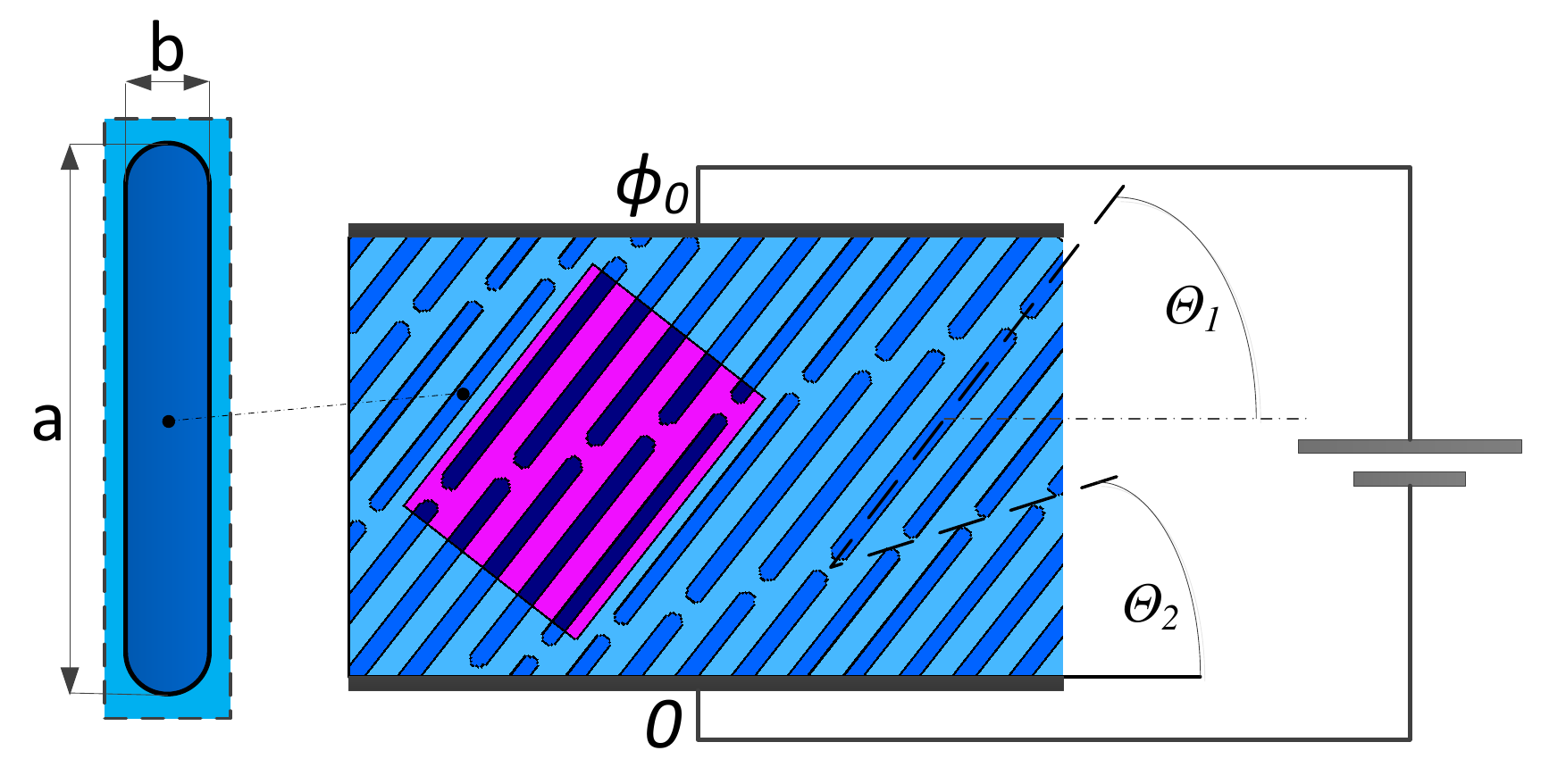}
\caption{A sketch of the periodic model.}\label{UC}
\end{figure}

A finite element model of a particulate periodic microstructure is shown in Fig.~\ref{UC}. 
The model consists of thin isotropic layers at an angle $\Theta_{2}$ between bulky stacks of elongated stiffer inclusions that are tilted at an angle $\Theta_{1}$. 
The highlighted inclined rectangle is a representative unit cell (RUC) of the model P5 with 5 inclusions.
A denser model, P10, with 10 inclusions in the RUC is analyzed too.
The aspect ratios $a/b$ are $16.35$ and $32.7$ for the models P5 and P10, respectively. 
The simulations were accomplished with appropriate in-plane periodic
displacement and potential boundary conditions.
The non-linear coupled problem was solved by application of COMSOL FE code.

Properties of the widely used 3M VHB-4910 scotch \cite{ohaletal08jap} were chosen for the soft phase in both, the idealized and the periodic, models. 
The properties for the stiffer high dielectric inclusions are comparable with those of Polyaniline (PANI) \cite{huanetal04aple}. 
The values of the physical constants are listed in Table~\ref{Pconstants}.

\begin{table}[b]
\caption{\label{Pconstants}%
Material constants}
\begin{ruledtabular}
\begin{tabular}{ccc}
 Phase &$\epsilon$&
\multicolumn{1}{c}{\textrm{$\mu~\mathrm{[MPa]}$}}\\
 \hline
 VHB 4910  &$6.5$
 & $0.2$  \\
Polyaniline  (PANI)   &$6500$  & $2700$ \\
\end{tabular}
\end{ruledtabular}
\end{table}

\begin{figure}[t]
\includegraphics[scale=0.68]{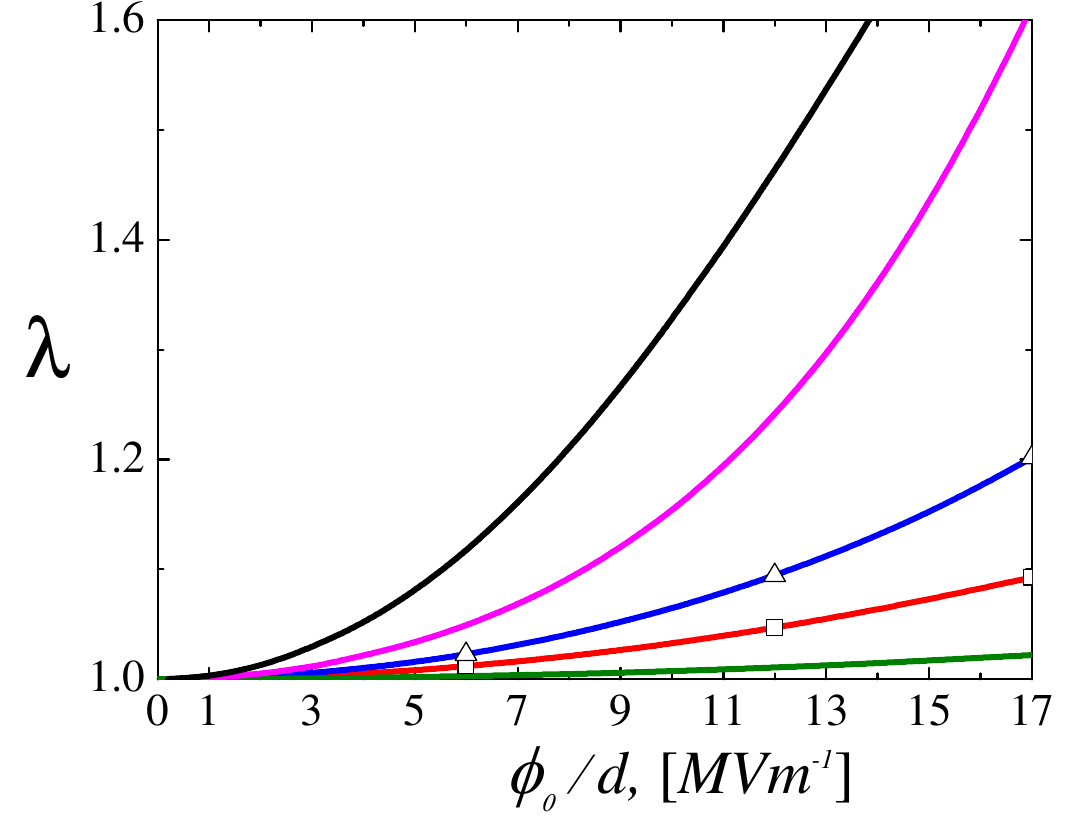}
\caption{Principal stretch ratio versus the electrostatic excitation. 
The black and magenta curves correspond to the idealized models I1 and I2.
The red and blue curves are the results of the simulations with models P5 and P10.
The green curve is the response of a homogeneous VHB-4910 actuator.}
\label{Lambda}
\end{figure}

Predictions for the principal actuation stretch as functions of the applied electric excitation 
are shown in Fig.~\ref{Lambda}.
The response of the homogeneous VHB-4910 actuator is determined via the relation \cite{ltd08phd}
\begin{equation}\label{HomForm}
    \lambda=\left(1-{\epsilon\epsilon_{0}}E_{0}^{2}/\mu\right)^{-1/4}.
\end{equation}
The optimized microstructure I1 displays the best response. 
Nonetheless, even the more realistic periodic microstructures exhibit a giant enhancement of the response relative to that of the homogeneous actuator. 
Specifically, at $17\textrm{MV/m}$ the periodic model P10 attains a $20\%$ strain in comparison with the $2\%$ strain of the homogeneous actuator.

\begin{figure}[t]
\includegraphics[scale=0.68]{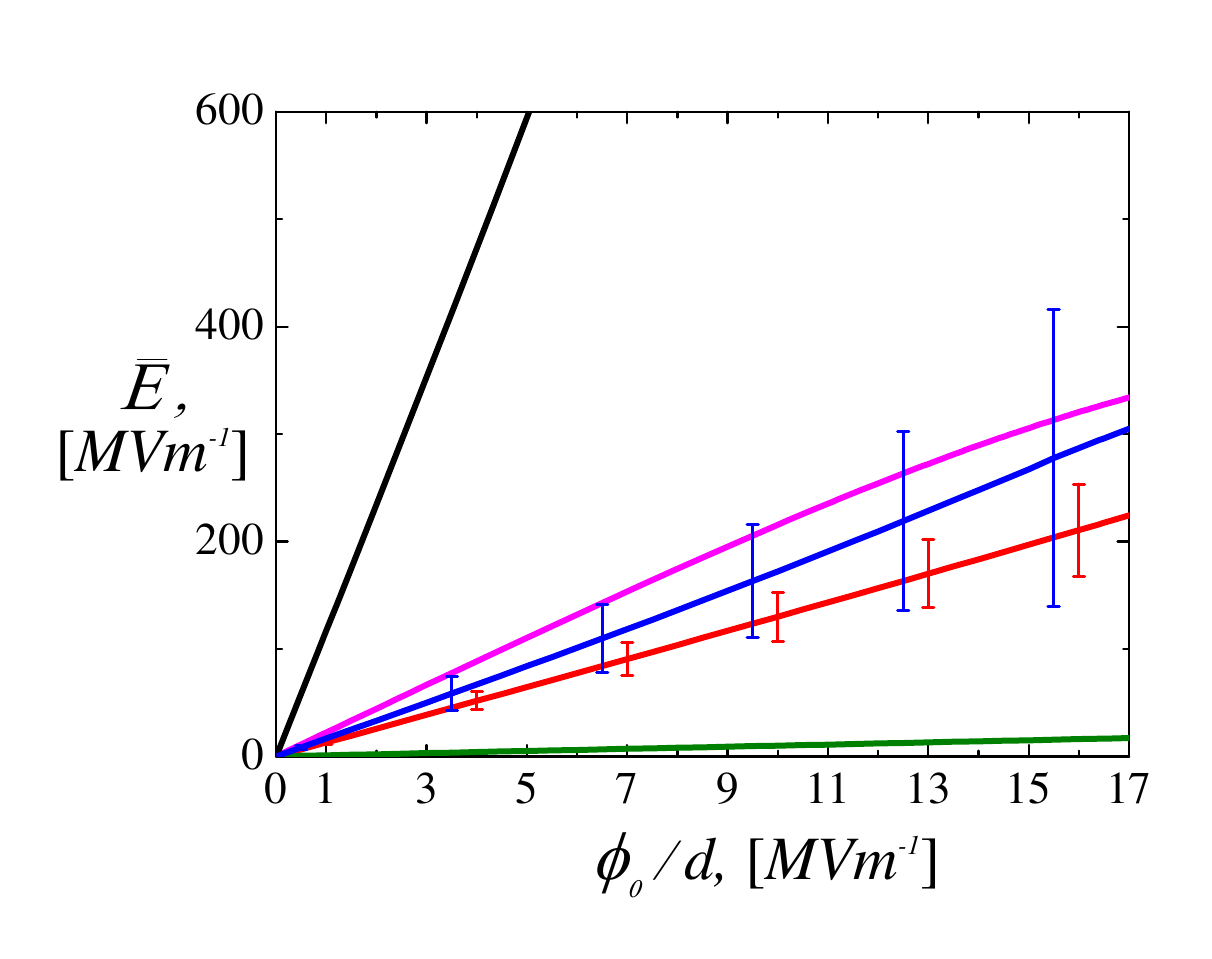}
\caption{The mean electric fields in the isotropic layers versus the electrostatic excitation. 
The black and magenta curves correspond to the idealized models I1 and I2.
The blue and red curves are the results of the FE models P10 and P5, and the error bars represent the standard deviation in the field.
The green curve depicts the field in a homogeneous VHB-4910 actuator.}
\label{Efield}
\end{figure}

The amplification of the electromechanical coupling is a tricky two-parts mechanism that highlights the roles of the fluctuations in the electric field and anisotropy.
Firstly, due to the continuity of the normal component of the electric displacement, 
the ratio between the mean electric fields in the anisotropic and the isotropic layers is inverse proportional to the ratio between their dielectric moduli.
Consequently, 
the electric field in the anisotropic layer is small,
and since the volume average of the electric field is equal to $E_{0}$, the magnitude of the electric field in the isotropic layers is proportional to the applied field divided by their volume fraction \citep{tianetal10submitted}.
In Fig.~\ref{Efield} the mean electric fields in the isotropic layers are shown.
Indeed, the intensities of these fields are approximately $E_{0}/\alpha^{(2)}$. 
As the number of stiff inclusions in the RUC increases, the curves for the periodic models approach the one for the idealized model I2.

The large electric field results in a large electrostatic stress
that tends to stretch the isotropic layer in the transverse direction in a manner reminiscent of the one shown in Fig.~\ref{actuator}.
Roughly speaking, due to the large fluctuations in the electric field, the isotropic layers act like ``micro-actuators''.
This brings us to the second stage of the amplification mechanism.
Since the anisotropic layers are made out of alternating stiff and compliant sublayers,
their compliant mode corresponds to a shear of the soft sublayers with rotation of the stiffer ones.
This mode amounts to an extension at $45^{\circ}$ to the sublayers plane, which is quite close to the angle $\Theta^{(2)}-\Theta^{(1)}$.
Thus, when the micro-actuators deform due to the intensive local electric field, they stretch the bulky anisotropic layers along their soft mode, and the actuator expands in a direction transverse to that of the electric excitation.

\begin{figure}[t]
\includegraphics[scale=0.21]{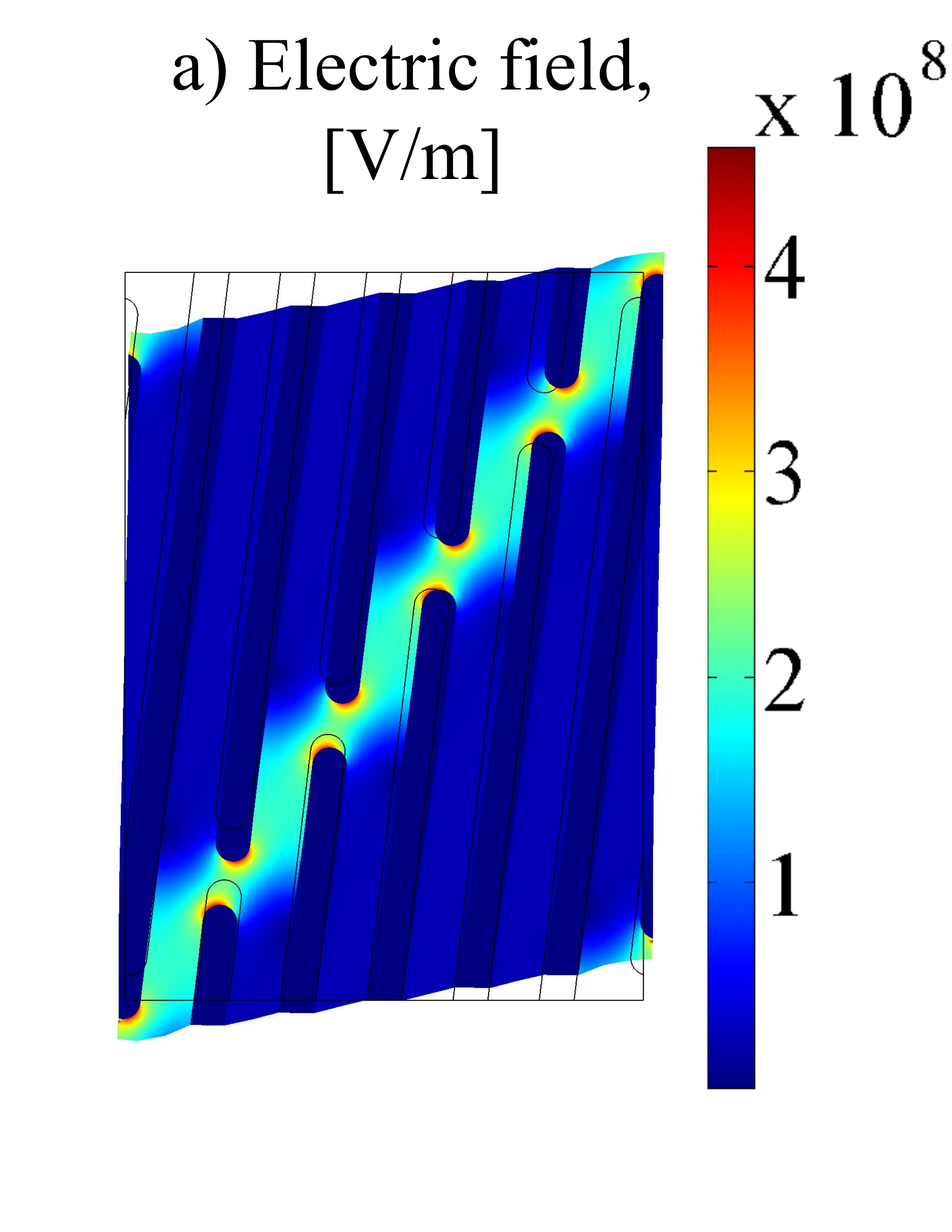}
\includegraphics[scale=0.21]{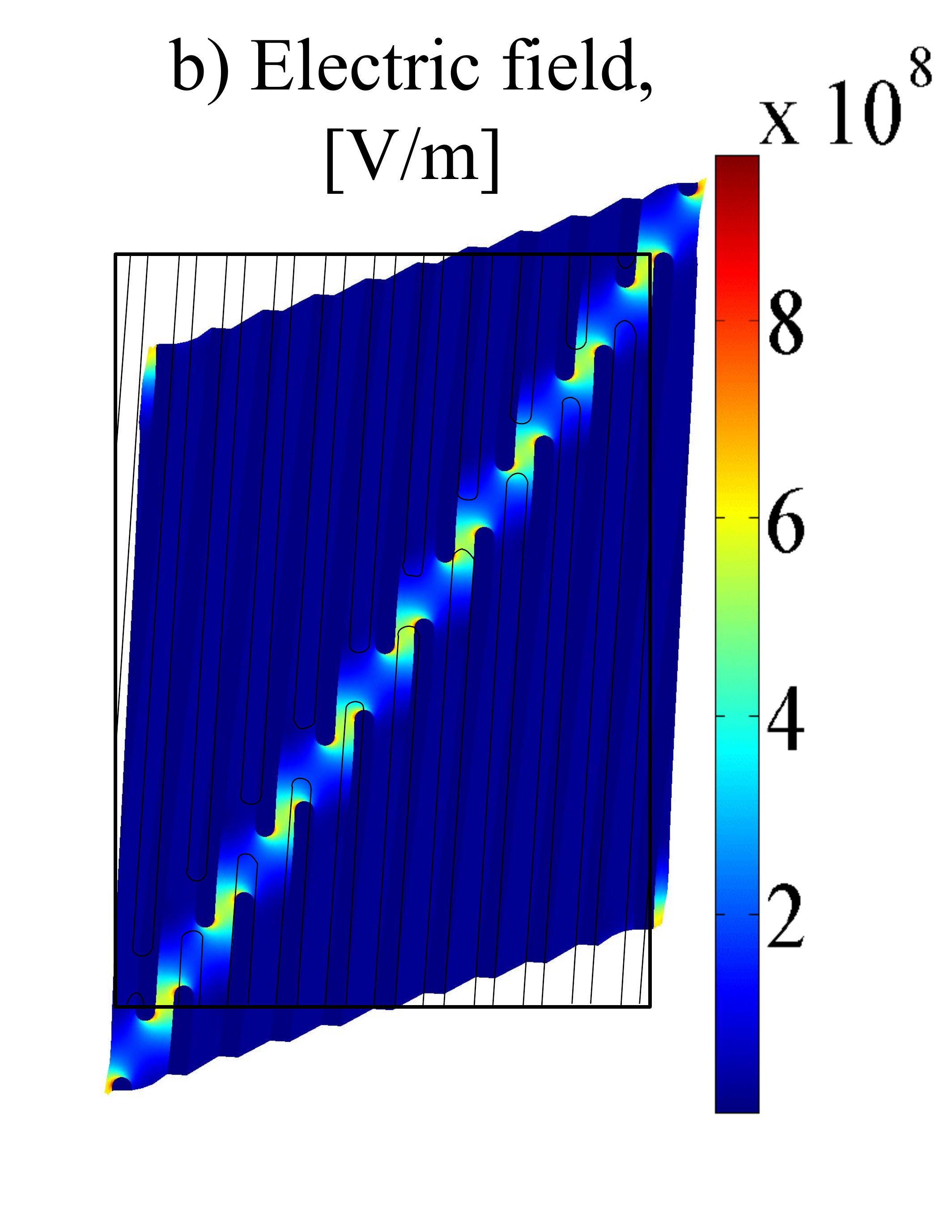}
\caption{The deformed state of the periodic RUCs and the distributions of the electric field in the models (a) P5 and (b) P10 at $E_{0}=17\textrm{MV/m}$.}
\label{mc}
\end{figure}

The amplification mechanism is illustrated in Fig.~\ref{mc}, where the intensification of the electric field in the micro-actuators, their resulting deformation perpendicular to the direction of the electric excitation, and the rotation of the elongated stiff inclusions can be appreciated.
We conclude noting that the proposed mechanism exhibits a \emph{ten-fold} enhancement of the electromechanical coupling, thus providing a method to overcome the primary obstacle in the development of EAP actuators.

%

\end{document}